\documentclass[prl,twocolumn,aps,superscriptaddress]{revtex4-2}

\usepackage{amsmath}
\usepackage{amssymb}
\usepackage{wasysym}
\usepackage{graphicx}
\usepackage{hyperref}
\usepackage{dsfont}
\usepackage{newtxtext}
\usepackage[varvw]{newtxmath}
\usepackage{mathtools}

\hypersetup{
    colorlinks,
    linkcolor={blue},
    citecolor={blue},
    urlcolor={blue}
}

\newcommand{\NCSO}{Na$_3$Co$_2$SbO$_6$}
\newcommand{\NCTO}{Na$_2$Co$_2$TeO$_6$}

\DeclareMathOperator{\diag}{diag}

\begin{document}

\title{Field-induced magnetic phases in the Kitaev candidate Na$_3$Co$_2$SbO$_6$}

\author{Kranthi Kumar Bestha}
\thanks{These two authors contributed equally to the work.}
\affiliation{Leibniz IFW Dresden, Institute of Solid State Research, 01069 Dresden, Germany}
\affiliation{Institut f\"ur Festk\"orper- und Materialphysik and W\"urzburg-Dresden Cluster of Excellence ct.qmat, Technische
Universit\"at Dresden, 01062 Dresden, Germany}
\author{Manaswini Sahoo}
\thanks{These two authors contributed equally to the work.}
\affiliation{Leibniz IFW Dresden, Institute of Solid State Research, 01069
Dresden, Germany}
\affiliation{Institut f\"ur Festk\"orper- und Materialphysik and W\"urzburg-Dresden Cluster of Excellence ct.qmat, Technische Universit\"at Dresden, 01062 Dresden, Germany}
\author{Niccol\`{o} Francini}
\affiliation{Institut f\"ur Theoretische Physik and W\"urzburg-Dresden Cluster of Excellence ct.qmat, Technische Universit\"at Dresden, 01062 Dresden, Germany}
\author{Robert Kluge}
\affiliation{Leibniz IFW Dresden, Institute of Solid State Research, 01069
Dresden, Germany}

\author{Ryan Morrow}
\affiliation{Leibniz IFW Dresden, Institute of Solid State Research, 01069
Dresden, Germany}
\author{Andrey Maljuk}
\affiliation{Leibniz IFW Dresden, Institute of Solid State Research, 01069
Dresden, Germany}
\author{Sabine Wurmehl}
\affiliation{Leibniz IFW Dresden, Institute of Solid State Research, 01069
Dresden, Germany}
\author{Sven Luther}
\affiliation{Hochfeld-Magnetlabor Dresden (HLD-EMFL), Helmholtz-Zentrum Dresden-Rossendorf, Dresden,
Germany}
\author{Yurii Skourski}
\affiliation{Hochfeld-Magnetlabor Dresden (HLD-EMFL), Helmholtz-Zentrum Dresden-Rossendorf, Dresden,
Germany}
\author{Hannes K\"uhne}
\affiliation{Hochfeld-Magnetlabor Dresden (HLD-EMFL), Helmholtz-Zentrum Dresden-Rossendorf, Dresden,
Germany}
\author{Swarnamayee Mishra}
\affiliation{Institut f\"ur Festk\"orper- und Materialphysik and W\"urzburg-Dresden Cluster of Excellence ct.qmat, Technische
Universit\"at Dresden, 01062 Dresden, Germany}

\author{Jochen Geck}
\affiliation{Institut f\"ur Festk\"orper- und Materialphysik and W\"urzburg-Dresden Cluster of Excellence ct.qmat, Technische
Universit\"at Dresden, 01062 Dresden, Germany}

\author{Manuel Brando}
\affiliation{Max Planck Institute for Chemical Physics of Solids, 01187 Dresden, Germany}

\author{Bernd B\"uchner}
\affiliation{Leibniz IFW Dresden, Institute of Solid State Research, 01069
Dresden, Germany}
\affiliation{Institut f\"ur Festk\"orper- und Materialphysik and W\"urzburg-Dresden Cluster of Excellence ct.qmat, Technische Universit\"at Dresden, 01062 Dresden, Germany}

\author{Laura T. Corredor}
\thanks{Present address: Faculty of Physics, Technical University of Dortmund, Otto-Hahn-Str. 4, D-44227 Dortmund, Germany}
\affiliation{Leibniz IFW Dresden, Institute of Solid State Research, 01069 Dresden, Germany}

\author{Lukas Janssen}
\email{lukas.janssen@tu-dresden.de}
\affiliation{Institut f\"ur Theoretische Physik and W\"urzburg-Dresden Cluster of Excellence ct.qmat, Technische Universit\"at Dresden, 01062 Dresden, Germany}

\author{Anja U.~B.~Wolter}
\email{a.wolter@ifw-dresden.de}
\affiliation{Leibniz IFW Dresden, Institute of Solid State Research, 01069
Dresden, Germany}

\date{\today}

\begin{abstract}
We report a rich anisotropic magnetic phase diagram of Na$_3$Co$_2$SbO$_6$, a previously proposed cobaltate Kitaev candidate, based on field- and temperature-dependent magnetization, specific heat, and magnetocaloric effect studies. At low temperatures, our experiments uncover a low-lying $j_{\textrm{eff}} = \frac{1}{2}$ state with an antiferromagnetic ground state and pronounced in-plane versus out-of-plane anisotropy. The experimentally identified magnetic phases are theoretically characterized through classical Monte Carlo simulations within an extended Kitaev-Heisenberg model with additional ring exchange interactions. The resulting phase diagram reveals a variety of exotic field-induced magnetic phases, including double-$\textbf{q}$, $\frac{1}{3}$-AFM, zigzag, and vortex phases.
\end{abstract}


\maketitle

\paragraph*{Introduction.}

In frustrated magnets, the quest for the elusive quantum spin liquid~\cite{savary17} has driven the discovery of diverse magnetic phases. The interplay of multiple exchange interactions can give rise to unconventional collinear and non-collinear magnetic orders with topological spin structures, which hold potential for device applications~\cite{Lohani2019}. In particular, materials hosting bond-dependent Kitaev exchange, in conjunction with other exchange interactions, can stabilize unique single-$\textbf{q}$ and multi-$\textbf{q}$ ground states, potentially leading to multiferroicity and the emergence of topological spin structures~\cite{janssen2016heisenbergkitaev, Li2023NiI2, kim2025higherorderskyrmioncrystalvan, Guan2024skyrmions, jin25}. Here, honeycomb cobalt oxides containing high-spin Co$^{2+}$ ions have attracted significant interest due to their ability to support strong Kitaev interactions alongside other symmetry-allowed exchange interactions~\cite{Liu2018, sano18}.

Cobalt compounds with $d^7$ electronic configuration exhibit unique characteristics, including an exchange process driven by $e_g$ electrons that cancels non-Kitaev interactions from $t_{2g}$-$t_{2g}$ channels, thereby enhancing the role of the Kitaev coupling. Notably, the $t_{2g}$-$e_g$ channel gives rise to a sizable Kitaev exchange, making cobaltates promising candidates to realize the various different states predicted in extended Kitaev-Heisenberg models~\cite{Liu2020, Kim2022review, rousochatzakis24, Winter2022}. The application of a magnetic field leads to additional exotic magnetic phases depending on the orientation and strength of the field~\cite{janssen2016heisenbergkitaev, janssen19, Gu2025multiq}. Among the honeycomb cobaltates, Na$_3$Co$_2$SbO$_6$ is expected to exhibit a weaker trigonal field, allowing for bond-dependent anisotropic interactions alongside comparable non-Kitaev interactions. These conditions make it a promising candidate for realizing the extended Kitaev-Heisenberg model~\cite{Mou2024RAMAN, Liu2023nonkitaev}.

Na$_3$Co$_2$SbO$_6$ has recently garnered significant interest~\cite{Songvilay2020NCTONCSO, viciu2007structure, stratan2019synthesis, yan2019magnetic, li2022giant, kim22, sanders22, Vavilova2023, miao2023persistent, Gu2024doubleq, hu24, mi2025}, particularly due to the predictions that tuning its trigonal crystal field via strain or pressure could bring it closer to the Kitaev spin liquid phase~\cite{Liu2020}. However, its antiferromagnetic (AFM) transition temperature is greatly influenced by the Na stoichiometry ($T_\mathrm{N}$ $\sim$ 4.3 - 8.3 K), and in the literature, the claimed ground state varies between zigzag and double-$\textbf{q}$~\cite{wong2016zig, Gu2024doubleq}. Notably, experimental strategies for distinguishing between single- and multi-$\textbf{q}$ orders, along with their implications for microscopic model construction, are currently a topic of intense debate. This discussion extends not only to cobaltates~\cite{li2022giant, Kruger2023NCTO, francini2024spinvestigial, francini2024ferrimagnetism, francini2025ferrimagnetismquantumfluctuationskitaev, jin25, Gu2025multiq} but also to other centrosymmetric compounds hosting three-dimensional topological spin structures~\cite{park23, Andriushin2025SrFeO3}. Furthermore, in Na$_3$Co$_2$SbO$_6$, the exact nature of the field-induced phases remains under intense scrutiny, with magnetic phase diagrams yet to be fully understood, particularly for the out-of-plane field direction. Thermodynamic studies on polycrystalline samples suggest a field-induced spin-liquid-like phase~\cite{Vavilova2023}, while AC composite magnetoelectric measurements, highly sensitive to spin fluctuations, reveal two field-induced tricritical points for in-plane magnetic fields, one of which appears to lie near a quantum critical point~\cite{mi2025}.

In this Letter, we investigate field-induced effects in high-quality single crystals of Na$_3$Co$_2$SbO$_6$ in both in-plane and out-of-plane directions, employing a variety of magnetic, thermodynamic, and theoretical methods. Our findings reveal a rich anisotropic magnetic phase diagram, highlighting novel field-induced magnetic phases that we characterize within an extended Kitaev-Heisenberg model with additional ring exchange interactions.

\paragraph*{Methods.}

Single crystals of Na$_{3}$Co$_{2}$SbO$_{6}$ were grown using the flux method and were fully characterized by powder and single-crystal X-ray diffraction (SCXRD) and energy-dispersive X-ray (EDX) analysis~\cite{supplemental}. DC magnetic-susceptibility and specific-heat measurements were carried out using a commercial Superconducting Quantum Interference Device (SQUID) magnetometer (MPMS3) and a Physical Property Measurement System (PPMS) from Quantum Design, respectively. Quartz and quartz paddles with two quartz half-cylinders holding the crystal vertically for out-of-plane measurements, were used for in-plane and out-of-plane magnetic measurements, respectively. Specific-heat measurements were conducted using the heat-pulse relaxation technique with a $^3$He insert in the PPMS.
%

%
Measurements of the bulk magnetization and magnetocaloric effect (MCE) were performed using pulsed magnetic fields at the Dresden High Magnetic Field Laboratory (HLD). Here, quasi-adiabatic conditions were ensured by the short duration of the magnetic-field pulse with a rise time of a few tens of ms, and by pumping the sample space to pressures below 1 mbar. The magnetization was recorded with a compensated pickup-coil magnetometer in a coaxial geometry. The background of each measurement was determined by a subsequent measurement with an empty magnetometer and otherwise identical conditions~\cite{Skourski2011}. The resulting pulsed-field magnetization curves were calibrated with data recorded in the MPMS. For the measurements of the MCE, the sample temperature was monitored with a RuOx-based thermometer glued directly to the sample, and corrected for magnetoresistive effects using a second sensor placed nearby. 

\begin{figure}[tb!]
\includegraphics[width=0.5\textwidth]{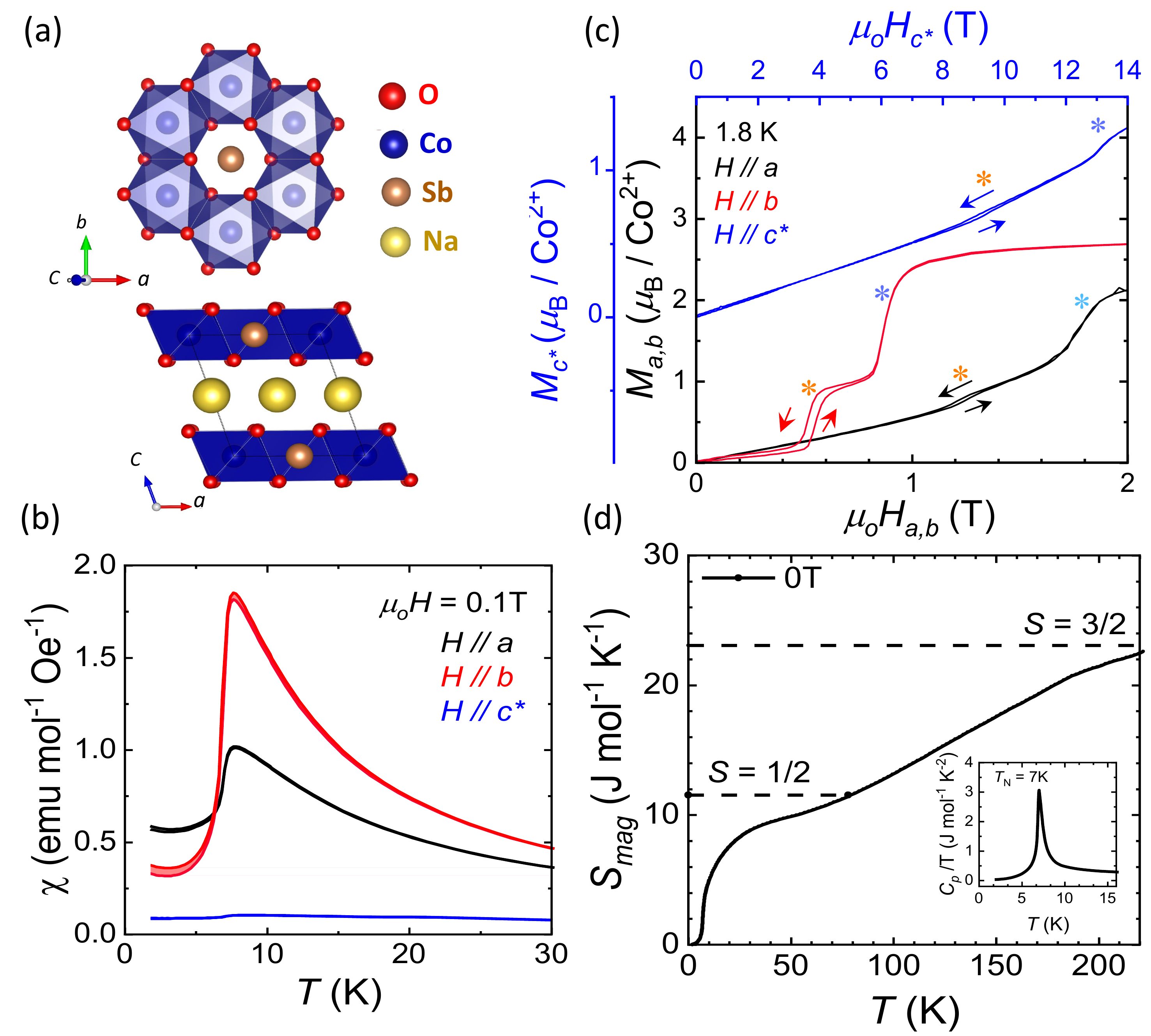}
\caption{%
(a)~Room-temperature crystal structure of Na$_3$Co$_2$SbO$_6$.
(b)~Magnetic susceptibility $\chi$ as a function of temperature $T$ at a field of $\mu_0 H = 0.1\,\mathrm{T}$ along the crystallographic $\mathbf{a}$, $\mathbf{b}$, and $\mathbf{c^*}  \perp \mathbf{a}, \mathbf{b}$ directions. The shaded areas highlight the bifurcation between field-cooled (upper curve) and zero-field-cooled (lower curve) magnetic susceptibility.
(c)~Isothermal magnetization $M$ as a function of field $H$ at $T = 1.8\,\mathrm{K}$ for $\mathbf H \parallel \mathbf {a}$ and $\mathbf H \parallel \mathbf {b}$ up to 2\,T, and at $T = 2\,\mathrm{K}$ for $\mathbf H \parallel \mathbf {c^*}$ up to 14\,T. Asterisks indicate the field-induced transitions (field-sweep up).
(d)~Magnetic entropy $S_\text{mag}$ as a function of temperature $T$ at zero field. The inset displays the zero-field heat capacity $C_p/T$.}
\label{fig:Fig1}
\end{figure}

\paragraph*{Experimental results.}

Figure~\ref{fig:Fig1}(a) illustrates the $C2/m$ crystal structure of Na$_3$Co$_2$SbO$_6$. The refinement of our SCXRD data resulted in lattice parameters $a=5.3855(3)\,\mathrm{\r{A}}$, $b = 9.2938(3)\,\mathrm{\r{A}}$, $c=5.6662(2)\,\mathrm{\r{A}}$, and $\beta = 108.6^\circ$ at room temperature. The lattice parameters indicate a tiny orthorhombic distortion, with $|\sqrt{3} a/b - 1| \lesssim 0.4\%$, consistent with previous reports~\cite{li2022giant}. The O-Co-O bond angles of the CoO$_6$ octahedra are $\sim 95^\circ$, indicating a significant compression along the interlayer stacking axis $\mathbf c^*$. Our refinement shows no detectable mixing of Co and Sb sites, and the EDX analysis reveals a homogeneous elemental distribution, indicative of the high quality of our single crystals.

The temperature dependence of the DC magnetic susceptibility $\chi$, with applied magnetic fields aligned along the in-plane directions $\mathbf{a}$ and $\mathbf{b}$, as well as the out-of-plane direction $\mathbf{c}^*$, is shown in Fig.~\ref{fig:Fig1}(b). A sharp decrease of the magnetic susceptibility is observed below $T_\mathrm{N} = 7\,\mathrm{K}$, indicating the onset of long-range antiferromagnetic (AFM) order. Here, $T_\mathrm{N}$ is defined as the inflection point in $\partial(\chi T)/\partial T$ according to Fisher's heat capacity definition~\cite{Fisher1962}, and it closely matches the N\'eel temperature reported previously~\cite{li2022giant}. Measurements on multiple single-crystal samples show the same $T_\mathrm{N}$, with a variation of less than 0.3\,K, indicating a homogeneous distribution of Na and near-ideal stoichiometry in our growth. The presence of magnetic order indicates the existence of non-Kitaev interactions, likely originating from the distorted octahedral environment [Co-O-Co bond angle of $\sim 92.7(4)^\circ$] and intraorbital hopping~\cite{Xiaoyu2023nonKitaev}. The tiny splitting between zero-field-cooled and field-cooled magnetic susceptibility below $T_\mathrm{N}$ indicates a weak uncompensated magnetic moment superimposed on the AFM order, consistent with previous observations~\cite{li2022giant}.

\begin{figure}[tb!]
\includegraphics[width=0.5\textwidth]{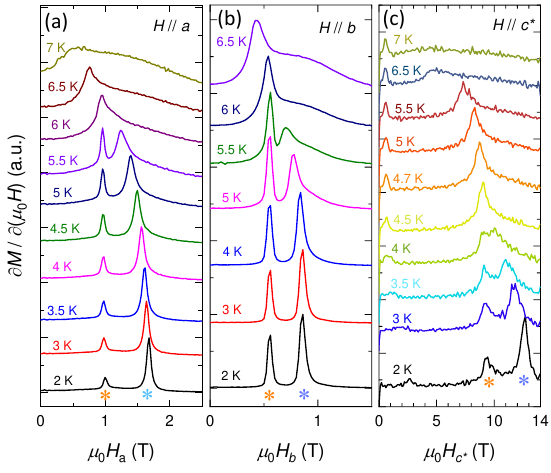}
\caption{%
(a)~Magnetic susceptibility $\partial M / \partial (\mu_0 H)$ for various fixed temperatures as a function of field $\mathbf H \parallel \mathbf a$ (field-sweep up) from SQUID magnetometry up to 2.5\,T.
Asterisks indicate the field-induced transitions.
(b)~Same as (a), but for $\mathbf H \parallel \mathbf b$ up to 1.5\,T.
(c)~Same as (a), but for $\mathbf H \parallel \mathbf c^*$ from vibrating sample magnetometry in a PPMS up to 14\,T.
}
\label{fig:Fig2}
\end{figure}

Figure~\ref{fig:Fig1}(c) shows the isothermal magnetization $M$ as a function of field at $T=1.8\,\mathrm{K}$ for $\mathbf H \parallel \mathbf {a}$ and $\mathbf H \parallel \mathbf {b}$ up to 2\,T, and at $T = 2\,\mathrm{K}$ for $\mathbf H \parallel \mathbf {c^*}$ up to 14\,T. Our $M(H)$ data reveal multiple distinct field-induced transitions, some of them showing strong hysteresis effects. Interestingly, field-induced transitions are also observed in the out-of-plane $\mathbf {c^*}$ direction, which was overlooked in previous studies~\cite{yan2019magnetic,li2022giant,wong2016zig}, highlighting the complex magnetic anisotropy in Na$_3$Co$_2$SbO$_6$. The observed critical field values vary by an order of magnitude depending on the direction, with the lowest field values found along the in-plane easy axis $\mathbf{b}$ and the highest values along the out-of-plane direction $\mathbf{c^*}$. The critical field values of the field-induced transitions for increasing magnetic fields (field-up sweeps) are obtained from the peak of the anomaly in the derivative curves $\partial M/\partial (\mu_0 H)$ at different temperatures, shown in Fig.~\ref{fig:Fig2}.
For fields along the in-plane direction $\mathbf a$ [Fig.~\ref{fig:Fig2}(a)] and $T=2$\,K, we observe two field-induced transitions at $\mu_0H_{a1} = 0.99 (50)$\,T, and $\mu_0H_{a2} = 1.68(12)$\,T. For fields along the in-plane direction $\mathbf{b}$ [Fig.~\ref{fig:Fig2}(b)] and $T = 2$\,K, we observe two transitions at $\mu_0 H_{b1} = 0.56(8)$\,T and $\mu_0 H_{b2} = 0.86 (5)$\,T, with the latter marking the complete suppression of the AFM order. The specific shape of $\partial M/\partial (\mu_0 H)$ near $H_{b2}$ for $T = 5$\,K might hint at short-range order in this region, in line with our specific-heat studies presented in the SM ~\cite{supplemental}.
Our results for the two in-plane directions are consistent with the literature~\cite{li2022giant}.
Note that the critical fields along $\textbf {a}$ exhibit large uncertainties due to a possible misorientation (in this work $\lesssim 10^\circ$) due to steep changes in the critical fields close to $\textbf {a}$, in contrast to the $\textbf{b}$ direction (see Supplemental Material (SM)~\cite{supplemental}).

For fields along the out-of-plane direction $\mathbf c^*$, the critical field strengths are significantly larger. 
For $T = 2$\,K, we observe two field-induced transitions at $\mu_0H_{c1} = 9.4 (4)$\,T and $\mu_0H_{c2}= 12.6 (8)$\,T (see Fig.~\ref{fig:Fig2}(c)) . The anomaly in the field range $0\leq\mu_0H\leq 2.6$\,T probably arises from short-range ferromagnetic chains formed above $T_\textrm{N}$~\cite{li2022giant} as this anomaly persists in the paramagnetic phase~\cite{supplemental}. 
Note that all transitions were consistently reproduced across different single-crystal samples, with slight variations in the critical field strengths ($\lesssim 0.4$\,T and $\lesssim 2$\,T for in and out-of-plane, respectively), likely due to minor differences in stoichiometry, disorder, or orientation. 
For all three directions, the $M(H)$ curves at 2\,K exhibit clear hysteresis at the lowest-field transition, indicating a first-order nature, which is further confirmed by our MCE measurements discussed below.

Overall, the isothermal magnetization reveals pronounced magnetic anisotropy.
From the quasi-saturated magnetizations, we estimate the anisotropic $g$ factors as $g_a = 6.2 \pm 0.3$, $g_b = 6.4 \pm 0.3$, and $g_{c^*} = 3.6 \pm 0.4$; see SM~\cite{supplemental} for details.
These values contrast with those of an undistorted octahedron, which is expected to exhibit an isotropic $g_\text{iso} = 4.33$ for a $j_\text{eff} = 1/2$ doublet ground state~\cite{abragam2012electron}.
In Na$_3$Co$_2$SbO$_6$, the distortion of the CoO$_6$ octahedra induces an admixture of $m = \pm 1/2$ components from the $j_{\text{eff}} = 3/2$ and $j_{\text{eff}} = 5/2$ multiplets with the $j_{\text{eff}} = 1/2$ state, giving rise to off-diagonal $\Gamma'$ interactions~\cite{rau14b}.
The significantly larger values of $g_a$ and $g_b$ (perpendicular to the trigonal field direction) compared to $g_{c^*}$ suggest that Co$^{2+}$ experiences a positive trigonal field.
The counterintuitive sign of the trigonal field inferred from the $g$-factor analysis, compared to the octahedral compression observed in SCXRD, suggests a significant ligand-field contribution from Sb ions, consistent with previous reports~\cite{Veenendaal2023XMCD}. Applying biaxial compressive strain in the $ab$ plane could thus be a viable approach to achieve the ideal, undistorted octahedral environment in Na$_3$Co$_2$SbO$_6$, similar to Cu$_3$Co$_2$SbO$_6$~\cite{Kim2024}.

The inset of Fig.~\ref{fig:Fig1}(d) shows the specific heat $C_p/T$ at zero field as a function of temperature. A clear peak at $T_\mathrm{N} = 7$\,K indicates the onset of long-range magnetic order.
No additional sharp anomalies were detected in our specific-heat measurements up to 220\,K (see SM~\cite{supplemental}). The magnetic contribution to the zero-field specific heat was extracted by subtracting the phonon contribution, obtained from the non-magnetic analogue compound Na$_3$Zn$_2$SbO$_6$. The corresponding zero-field magnetic entropy $S_\text{mag}$ = $\int\mathrm{d} T\, C_\text{mag}/T $ is shown in Fig.~\ref{fig:Fig1}(d). 
For temperatures $T \lesssim 50$\,K, the entropy reaches a plateau at around $S_\text{mag}^\text{theory} = 2R \ln 2 = 11.5$\,J/mol/K, consistent with the expectation for a system containing two magnetic Co$^{2+}$ ions with $j_{\text{eff}} = 1/2$.
The accumulated entropy up to $T_\mathrm{N} = 7$\,K represents only $\sim15\%$ of $S_\text{mag}^\text{theory}$, indicating frustrated quasi-2D short-range spin correlations reaching up to $\sim 50$\,K.
The additional magnetic entropy release above $\sim 80$\,K arises from excitations to $j_\text{eff} = 3/2$ states, consistent with the observation of an additional mode in Raman spectroscopy~\cite{Ponosov2024}.
Our modified Curie-Weiss fit of the magnetic susceptibility for $\mathbf H \parallel \mathbf {c^*}$ supports this conclusion and indicates an excitation gap of $\Delta E_{\frac{1}{2}\to \frac{3}{2}} = 247(3)$\,K (see SM~\cite{supplemental}).

\begin{figure}[tb!]
\includegraphics[width=0.5\textwidth]{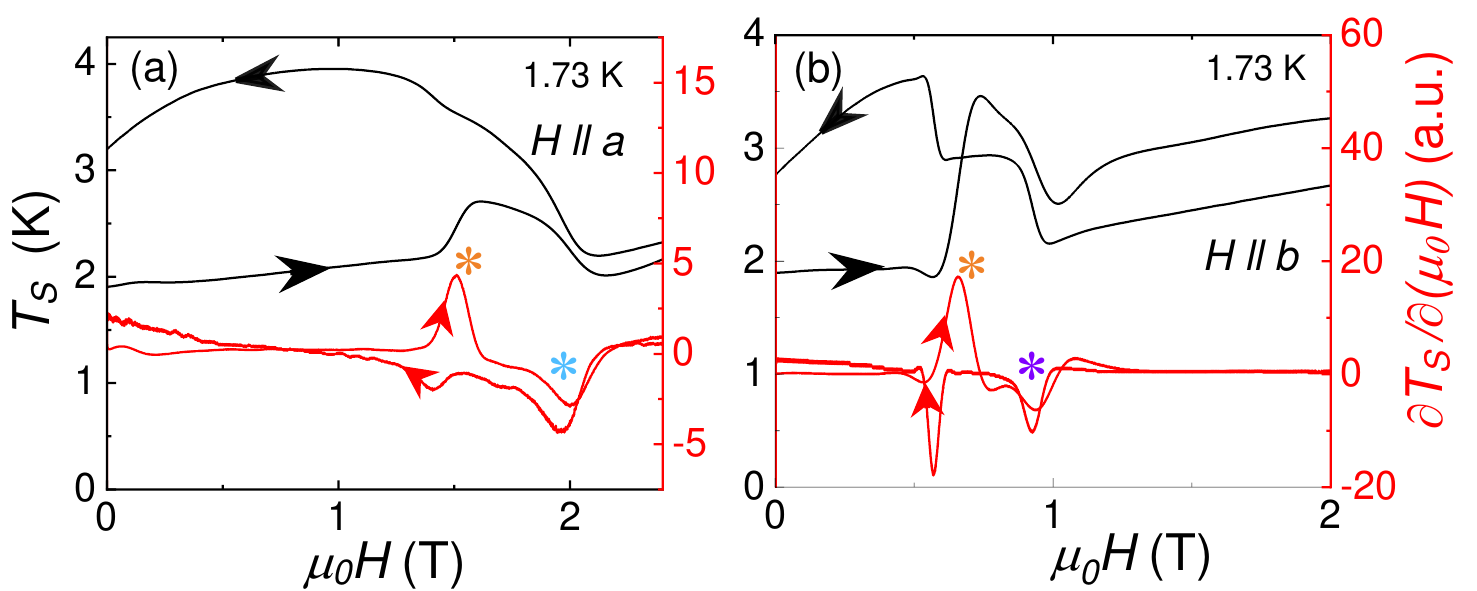}
\caption{%
(a)~Sample temperature $T_\text{S}$ (black) and its derivative $\partial T_\text{S}/\partial (\mu_0 H)$ (red) as a function of field $\mathbf H  \parallel \mathbf a$ from low-temperature MCE measurements in pulsed fields under quasi-adiabatic conditions. Arrows indicate the field-up and field-down sweeps, while asterisks mark the field-induced transitions.
(b)~Same as (a), but for $\mathbf H  \parallel \mathbf b$.
%
%
}
 \label{fig:Fig3}
\end{figure}

Our MCE measurements under quasi-adiabatic conditions in pulsed magnetic fields, shown in Fig.~\ref{fig:Fig3} for both field sweep-up and sweep-down, provide additional insights into the nature of the field-induced magnetic phases in Na$_3$Co$_2$SbO$_6$. For these studies, a different set of single crystals was used, resulting in slightly different critical field values.
In general, the critical field values are determined from the minima in $\partial T_\text{S}/\partial H$, where $T_\text{S}$ refers to the sample temperature. This definition is consistent with an increase in spin disorder near the phase transition, which forces a drop in thermal entropy to compensate. For a first-order transition involving latent heat, however, the quasi-adiabatic condition may be violated, resulting in maxima in $\partial T_\text{S}/\partial H$ due to an abrupt increase in the sample temperature. 
For the $a$ direction, Fig.~\ref{fig:Fig3}(a), during the field-up sweep, the sample temperature increases monotonically until the critical field $\mu_0H_{a1} \sim 1.5$\,T, indicating a decrease in magnetic entropy.
At the critical field $H_{a1}$, a sudden increase in the sample temperature during the field-up sweep marks the release of latent heat. Conversely, the field-down sweep exhibits a further temperature increase below $H_{a1}$, leading to a negative slope $\partial T_\text{S}/\partial H < 0$, confirming the first-order nature of the phase transition. 
%
%
The strong irreversibility between the sweep-up and sweep-down processes may be attributed to AFM domain dynamics and the complex field evolution of the double-$\mathbf q$ spin structure~\cite{li2022giant}, with a loss mechanism driven by spin fluctuations.
A local minimum in $T_\text{S}$ marks the critical field $H_{a2}$. This behavior is consistently observed in both field-up and field-down sweeps, suggesting the second-order nature of this transition. Above $H_{a2}$, the steady increase in $T_\text{S}$ reflects the opening of a spin-excitation gap in the field-polarized state, compensating for the drop in spin entropy. This is in line with our specific-heat measurements for fields above $H_{a2}$ (see SM~\cite{supplemental}).
A similar behavior has also been observed for the $b$ direction, where the field-induced transitions at the two critical fields are marked by a sudden increase and a local minimum in the sample temperature, respectively, corresponding to first-order ($H_{b1}$) and second-order ($H_{b2}$) phase transitions. Contrary to the $a$ direction, a strong hysteresis is observed above $H_{b2}$ in the field-polarized state, suggesting some irreversibility arising from mechanisms beyond simple magnetic-domain effects.
For the $c^*$ axis, the critical field $H_{c2}$ is identified for field-up sweeps in our MCE study; see SM~\cite{supplemental}.

\begin{figure*}[tb!]
    \includegraphics[width=1\textwidth]{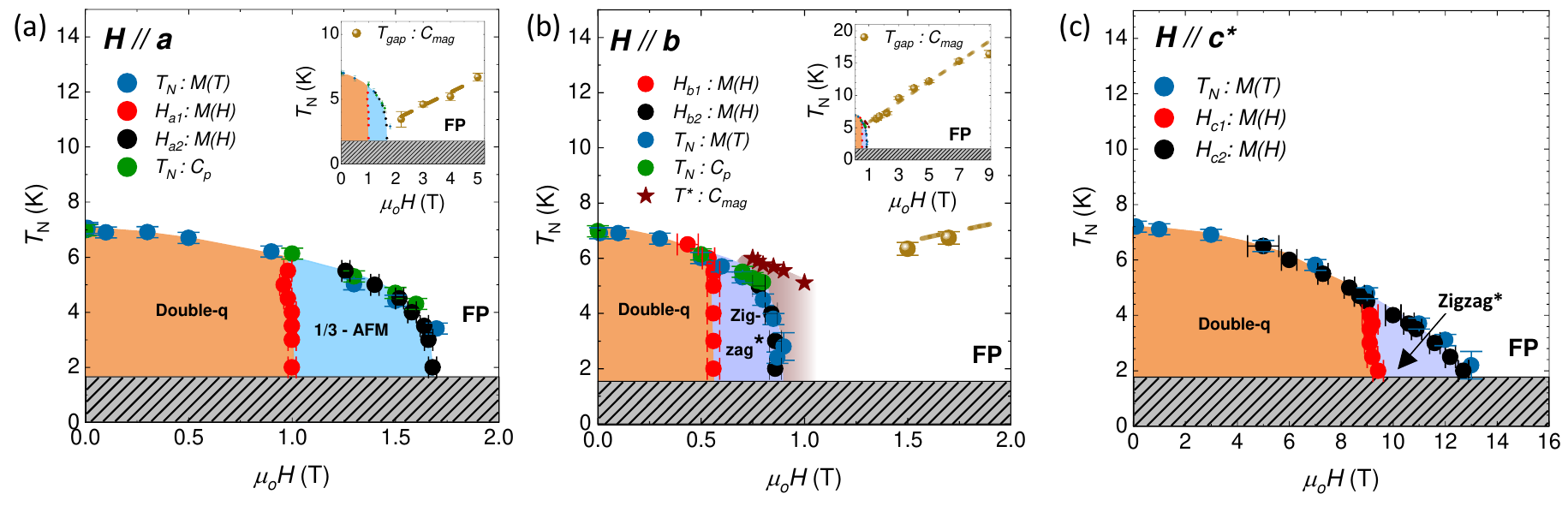}\\[-1\baselineskip]
    \includegraphics[width=1\textwidth]{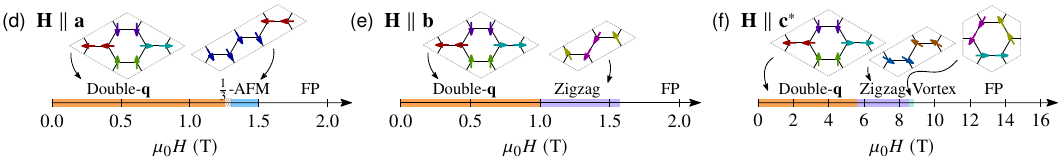}
    \caption{%
    Top: Na$_3$Co$_2$SbO$_6$ experimental magnetic phase diagram for 
    (a)~$\mathbf H \parallel \mathbf{a}$, 
    (b)~$\mathbf H \parallel \mathbf{b}$, 
    and 
    (c)~$\mathbf H \parallel \mathbf{c^*}$. Note that the phase boundaries correspond to the critical fields observed for field-up sweeps.
    The field-induced phases predicted by theory are marked with an asterisk, while those without an asterisk are based on previous neutron diffraction experiments~\cite{li2022giant}.
    Insets illustrate the spin-excitation gap in the field-polarized (FP) state. 
    %
    %
    %
    Bottom: Ground-state phase diagram of extended Kitaev-Heisenberg model for
    (d)~$\mathbf H \parallel \mathbf {a}$,
    (e)~$\mathbf H \parallel \mathbf {b}$,
    and
    (f)~$\mathbf H \parallel \mathbf {c^*}$.
    Insets indicate spin directions projected onto the $ab$ plane in the different phases.
    %
    %
    }
 \label{fig:phase diagram}
\end{figure*}

The anisotropic field and temperature evolution of the magnetic ground state of Na$_3$Co$_2$SbO$_6$ as determined through magnetic and specific heat measurements is summarized in Figs.~\ref{fig:phase diagram}(a--c) for the main crystallographic directions $\mathbf a$, $\mathbf b$, and $\mathbf c^*$. Along the $b$ direction, in the field range $0.7 \leq \mu_0H \leq 0.85$\,T, just after the suppression of magnetic long-range order, a short-range magnetic order emerges, indicated by dark red stars in Fig.~\ref{fig:phase diagram}(b). Increasing the magnetic field beyond the suppression of AFM order leads to the opening of a spin-excitation gap, as revealed by our specific-heat measurements, see insets of Figs.~\ref{fig:phase diagram}(a,b) and the SM~\cite{supplemental}. We note that it is beyond the scope of this work to speculate on the existence of a possible gapless regime between the full suppression of long-range AFM order and the opening of the spin gap. However, if such a regime exists, its width would not exceed $\Delta \mu_0H \sim 0.5$\,T.

\paragraph*{Modeling.}

In order to characterize the experimentally observed field-induced phases theoretically, we use classical Monte Carlo simulations~\cite{francini2024spinvestigial, francini2024ferrimagnetism} to study an extended Kitaev-Heisenberg model, defined by the Hamiltonian
\begin{align}
\label{eq:model}
\mathcal H & = 
\sum_{\langle ij \rangle_{\gamma}} 
\Bigl[ J \mathbf{S}_i \cdot \mathbf{S}_j + K S_{i}^{\gamma}S_{j}^{\gamma}
+ \Gamma(S_{i}^{\alpha} S_{j}^{\beta} + S_{i}^{\beta}S_{j}^{\alpha})
\nonumber \displaybreak[0] \\ & \quad
+ \Gamma'(S_{i}^{\gamma}S_{j}^{\alpha} + S_{i}^{\alpha}S_{j}^{\gamma} + S_{i}^{\gamma}S_{j}^{\beta} + S_{i}^{\beta}S_{j}^{\gamma}) \Bigr] 
+ \mathcal H_{\hexagon}^{C_2}
\nonumber \displaybreak[0] \\ & \quad
- \mu_0 \mathbf H \cdot \sum_i \mu_\mathrm{B} g \mathbf S_i\,,
\end{align}
with nearest-neighbor bilinear exchange couplings $J$, $K$, $\Gamma$, and $\Gamma'$, a nonbilinear ring exchange $\mathcal H_{\hexagon}^{C_2}$, a Zeeman coupling of the effective moment $\mathbf M = \sum_i \mu_\mathrm{B} g \mathbf S_i$ to the magnetic field $\mu_0 \mathbf H$, and a diagonal $g$ tensor $g = \diag(g_a, g_b, g_{c^*})$, with $g_{a,b,c^*}$ taken from the high-field magnetization measurements discussed above.
Ring exchange interactions have recently been argued to be of relevance in the sister compound \NCTO~\cite{Kruger2023NCTO, francini2024ferrimagnetism, francini2024spinvestigial, francini2025ferrimagnetismquantumfluctuationskitaev}. Here, we assume a form consistent with the $C_2^*$ of the monoclinic lattice~\cite{li2022giant}, $\mathcal H_{\hexagon}^{C_2} = J_{\hexagon} \sum_{\langle ijklmn \rangle} 
[(\mathbf{S}_i \cdot \mathbf{S}_l)(\mathbf{S}_j \cdot \mathbf{S}_k)(\mathbf{S}_m \cdot \mathbf{S}_n) + 
(\mathbf{S}_i \cdot \mathbf{S}_k)(\mathbf{S}_j \cdot \mathbf{S}_m)(\mathbf{S}_l \cdot \mathbf{S}_n)
+ (ikm) \leftrightarrow (jnl)]$, 
with the summation over elemental plaquettes involving sites
$(ijklmn)$ in counter-clockwise order, and $(ikm) \leftrightarrow (jnl)$ denotes terms arising from the first two terms by an exchange of lattice indices, corresponding to symmetrization with respect to the $C_2^*$ symmetry.
Details of the simulations are given in the SM~\cite{supplemental}.
We choose $(J,K,\Gamma,\Gamma',J_{\hexagon}) = A (-1/9,-2/3,3,-4/9,-3/20)$, where $A = 0.5\ \text{meV}$ corresponds to the overall energy scale.
These parameters have been chosen to recover
(1)~the zero-field double-$\mathbf q$ ground state found in neutron diffraction experiments~\cite{li2022giant}
and
(2)~the sizable in-plane vs.\ out-of-plane magnetic anisotropy manifested in Figs.~\ref{fig:phase diagram}(a)--(c).
Remarkably, as illustrated in the ground-state phase diagrams shown in Figs.~\ref{fig:phase diagram}(d)--(f), the model constructed this way automatically reproduces several further features of \NCSO\ found in our measurements and previous literature results, namely
(3)~the presence of field-induced intermediate ordered phases for all three field directions
and
(4)~an evolution of the spin structure factor with Bragg peaks at momenta $\mathbf q = \mathbf M$ (double-$\mathbf q$ phase) for small fields, shifting to $\mathbf q = (2/3) \mathbf M$ for fields $\mathbf H \parallel \mathbf a$ in the phase below the high-field transition (1/3-AFM phase), in agreement with the neutron diffraction results~\cite{li2022giant}.
The agreement between theory and experiment allows us to predict the nature of the field-induced phases that have previously not been characterized in the literature. In particular, our modeling suggests that the field-induced intermediate ordered phases observed for $\mathbf H \parallel \mathbf b$ and $\mathbf H \parallel \mathbf c^*$ realize canted zigzag order, in contrast to the field-induced 1/3-AFM order observed for $\mathbf H \parallel \mathbf a$. The conclusion that the field-induced orders for $\mathbf H \parallel \mathbf a$ and $\mathbf H \parallel \mathbf b$ are not adiabatically connected is supported by magnetization measurements as a function of the in-plane field angle, presented in the SM~\cite{supplemental}.
The new phases predicted from the modeling are marked with an asterisk in Figs.~\ref{fig:phase diagram}(a)--(c).
Corresponding real-space spin configurations are sketched in the insets of Figs.~\ref{fig:phase diagram}(d)--(f) and further characterization is presented in the SM~\cite{supplemental}.

\paragraph*{Conclusions.}

Our study on bulk single crystals of Na$_3$Co$_2$SbO$_6$ reveals an AFM order below $T_\mathrm{N}$ with strong in-plane versus out-of-plane anisotropy. Our results establish a $j_{\textrm{eff}} = \frac{1}{2}$ ground state that is well-separated from the excited $j_{\textrm{eff}} = \frac{3}{2}$ state at low temperatures. The anisotropic magnetic phase diagram was constructed from temperature- and field-dependent specific-heat and magnetization measurements along all three crystallographic axes, revealing multiple field-induced phases. The nature of the transitions has been determined through quasi-adiabatic magnetocaloric effect studies. Using classical Monte Carlo simulations within an extended $K$-$J$-$\Gamma$-$\Gamma'$ Kitaev-Heisenberg model that incorporates $C_2$ ring exchange interactions, we have characterized the experimentally observed field-induced phases. Our results call for neutron diffraction studies to validate the field-induced canted zigzag orders predicted for fields along the $b$ and $c^*$ directions. Measuring the excitation spectrum using inelastic neutron scattering in the field-polarized phase could help to determine the most suitable effective spin model for Na$_3$Co$_2$SbO$_6$, similar to recent work on BaCo$_2$(AsO$_4$)$_2$~\cite{maksimov25}.

\begin{acknowledgments} 
\paragraph{Acknowledgments.}
Technical support in magnetization measurements by F.~Boden and S.~Ga\ss{} is gratefully acknowledged.
We thank
A.~Alfonsov,
A.~L.~Chernyshev, 
L.~Hozoi,
O.~Janson, 
V.~Kataev,
V.~Kocsis,
H.~J.~Grafe, 
and 
P.~A.~Maksimov
for insightful discussions.
Crystallographic visualizations were generated using VESTA~\cite{momma11}.
This work has been funded by the Deutsche Forschungsgemeinschaft (DFG) under
Project No.~247310070 (SFB~1143, Projects A07, B01, \& C01),
%
%
Project No.~390858490 (W\"urzburg-Dresden Cluster of Excellence on Complexity and Topology in Quantum Matter – ct.qmat, EXC~2147), and
Project No.~411750675 (Emmy Noether program, JA2306/4-1).
This work was supported by HLD-HZDR, member of the European Magnetic Field Laboratory (EMFL).
The authors gratefully acknowledge the computing time made available to them on the high-performance computer Barnard at the NHR Center of TU Dresden. This center is jointly supported by the Federal Ministry of Education and Research and the state governments participating in the National High-Performance Computing (NHR) joint funding program~\cite{nhr-alliance}.

\end{acknowledgments}

\bibliographystyle{longapsrev4-2}
\bibliography{NCSO}
\end{document}


\title{Supplemental Material for\\
``Field-induced magnetic phases in the Kitaev candidate Na$_3$Co$_2$SbO$_6$''}

\author{Kranthi Kumar Bestha}
\thanks{These two authors contributed equally to the work.}
\affiliation{Leibniz IFW Dresden, Institute of Solid State Research, 01069 Dresden, Germany}
\affiliation{Institut f\"ur Festk\"orper- und Materialphysik and W\"urzburg-Dresden Cluster of Excellence ct.qmat, Technische
Universit\"at Dresden, 01062 Dresden, Germany}
\author{Manaswini Sahoo}
\thanks{These two authors contributed equally to the work.}
\affiliation{Leibniz IFW Dresden, Institute of Solid State Research, 01069
Dresden, Germany}
\affiliation{Institut f\"ur Festk\"orper- und Materialphysik and W\"urzburg-Dresden Cluster of Excellence ct.qmat, Technische Universit\"at Dresden, 01062 Dresden, Germany}
\author{Niccol\`{o} Francini}
\affiliation{Institut f\"ur Theoretische Physik and W\"urzburg-Dresden Cluster of Excellence ct.qmat, Technische Universit\"at Dresden, 01062 Dresden, Germany}
\author{Robert Kluge}
\affiliation{Leibniz IFW Dresden, Institute of Solid State Research, 01069
Dresden, Germany}

\author{Ryan Morrow}
\affiliation{Leibniz IFW Dresden, Institute of Solid State Research, 01069
Dresden, Germany}
\author{Andrey Maljuk}
\affiliation{Leibniz IFW Dresden, Institute of Solid State Research, 01069
Dresden, Germany}
\author{Sabine Wurmehl}
\affiliation{Leibniz IFW Dresden, Institute of Solid State Research, 01069
Dresden, Germany}
\author{Sven Luther}
\affiliation{Hochfeld-Magnetlabor Dresden (HLD-EMFL), Helmholtz-Zentrum Dresden-Rossendorf, Dresden,
Germany}
\author{Yurii Skourski}
\affiliation{Hochfeld-Magnetlabor Dresden (HLD-EMFL), Helmholtz-Zentrum Dresden-Rossendorf, Dresden,
Germany}
\author{Hannes K\"uhne}
\affiliation{Hochfeld-Magnetlabor Dresden (HLD-EMFL), Helmholtz-Zentrum Dresden-Rossendorf, Dresden,
Germany}
\author{Swarnamayee Mishra}
\affiliation{Institut f\"ur Festk\"orper- und Materialphysik and W\"urzburg-Dresden Cluster of Excellence ct.qmat, Technische
Universit\"at Dresden, 01062 Dresden, Germany}

\author{Jochen Geck}
\affiliation{Institut f\"ur Festk\"orper- und Materialphysik and W\"urzburg-Dresden Cluster of Excellence ct.qmat, Technische
Universit\"at Dresden, 01062 Dresden, Germany}

\author{Manuel Brando}
\affiliation{Max Planck Institute for Chemical Physics of Solids, 01187 Dresden, Germany}

\author{Bernd B\"uchner}
\affiliation{Leibniz IFW Dresden, Institute of Solid State Research, 01069
Dresden, Germany}
\affiliation{Institut f\"ur Festk\"orper- und Materialphysik and W\"urzburg-Dresden Cluster of Excellence ct.qmat, Technische Universit\"at Dresden, 01062 Dresden, Germany}

\author{Laura T. Corredor}
\thanks{Present address: Faculty of Physics, Technical University of Dortmund, Otto-Hahn-Str. 4, D-44227 Dortmund, Germany}
\affiliation{Leibniz IFW Dresden, Institute of Solid State Research, 01069 Dresden, Germany}

\author{Lukas Janssen}
\email{lukas.janssen@tu-dresden.de}
\affiliation{Institut f\"ur Theoretische Physik and W\"urzburg-Dresden Cluster of Excellence ct.qmat, Technische Universit\"at Dresden, 01062 Dresden, Germany}

\author{Anja U.~B.~Wolter}
\email{a.wolter@ifw-dresden.de}
\affiliation{Leibniz IFW Dresden, Institute of Solid State Research, 01069
Dresden, Germany}

\begin{abstract}
The Supplemental Material contains additional information on
(i)~the crystal growth and the structural analysis of our single crystals, 
(ii)~DC magnetization measurements in static and pulsed fields, along with further analysis, and further magnetocaloric effect (MCE) data,
(iii)~specific-heat measurements of Na$_3$Co$_2$SbO$_6$ and its nonmagnetic analogue compound Na$_3$Zn$_2$SbO$_6$, 
and
(iv)~details of the Monte Carlo simulations, along with further characterization of the field-induced phases.
\end{abstract}

\date{\today}

\maketitle

\section{Crystal growth and characterization}

This supplemental section outlines the crystal growth process and structural analysis of our single crystals.

\subsection{Crystal growth}

The as-sintered Na$_3$Co$_2$SbO$_6$ powder was added to the (Na$_2$O + Sb$_2$O$_3$) flux in the 1.0:0.5:2.0 mol ratio in the glove box. The powder mixture was put in a Pt crucible, tightly covered by a Pt lid. The whole setup was heated in the box furnace up to 1400\,\degree C, held at this temperature for 6\,h, and then cooled at 2\,\degree C/h up to 1300\,\degree C. At this temperature, the box furnace was turned off and cooled to room temperature. The bulky Na$_3$Co$_2$SbO$_6$ single crystals were mechanically detached from the Pt crucible. The residual flux was dissolved in 1M NaOH. The as-grown samples were characterized by scanning electron microscopy with energy dispersive X-ray spectroscopy (EDX), powder X-ray diffraction, and Laue patterns. No Pt contamination was detected by EDX analysis. No impurity phases were found in flux-grown samples by powder XRD measurements.

\subsection{Single-crystal XRD}

\begin{figure*}[tb!]
\includegraphics[width=1\textwidth]{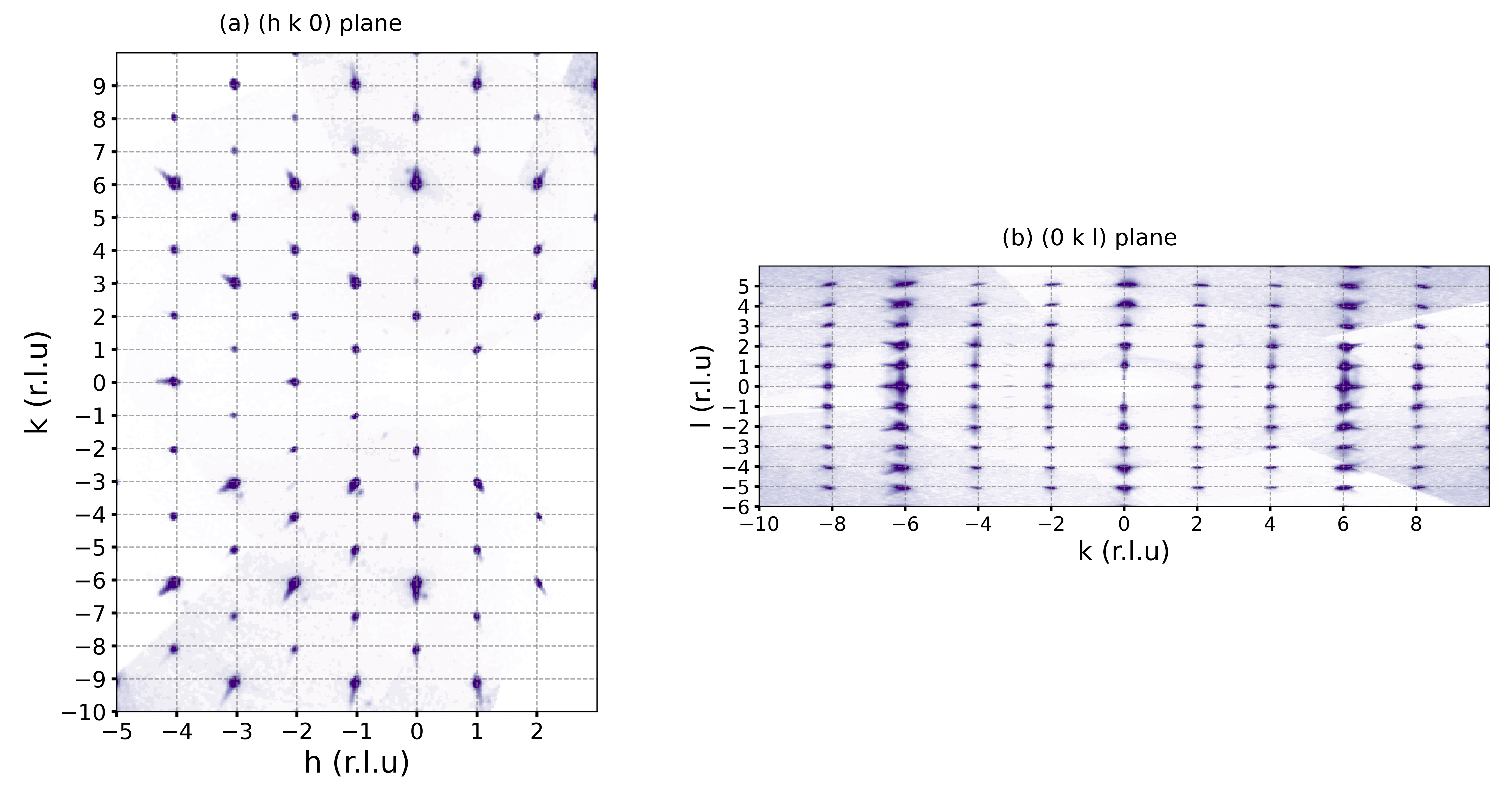}
\caption{%
(a)~Single-crystal X-ray diffraction pattern of Na$_3$Co$_2$SbO$_6$ at room temperature along the $(h k 0)$ plane in reciprocal space.
(b)~Same as (a), but along the $(0 k l)$ plane.}
\label{fig:xray}
\end{figure*}

\begin{table}[b!]
\caption{Crystallographic data obtained from Rietveld refinement of single-crystal X-ray diffraction data of Na$_3$Co$_2$SbO$_6$ at room temperature. Space group $C2/m$ (No.\ 12); $a = 5.3855(3)$\AA, ${b} = 9.2938(3)$\AA, ${c} = 5.6662(2)$\AA, $\beta = 108.558(6)^{\circ}$, ${V} = 268.86(2){\mathrm\AA}^3$.}.
\label{tab:xray-refinement}
\begin{ruledtabular}
\begin{tabular}{ccccccc}
\textbf{Atom} & \textbf{Site} & \bf\textit{x}  & \bf\textit{y} & \bf\textit{z} & {\bf\textit{U\textsubscript{\rm \bf iso}} (\AA$^2$)} & \textbf{Occup.}\\
\hline
Na1 & 2\textit{c} & 1.0000 & 0.0000 & 0.5000 & 0.0072(16) & 1 \\
Na2 & 4\textit{h} & 0.5000 & 0.1716(5) & 0.5000 & 0.0088(15) & 1 \\
Co  & 4\textit{g} & 0.5000 & 0.3333 & 0.0000 & 0.0025(5) & 1 \\
Sb  & 2\textit{2} & 0.5000 & 0.0000 & 0.0000 & 0.0080(4) & 1 \\
O1  & 4\textit{4} & 0.2510(19) & 0.0000 & 0.2092(17) & 0.0066(15) & 1 \\ 
O2  & 8\textit{8} & 0.7255(17) & 0.1603(5) & 0.2077(15) & 0.0059(13) & 1 \\
\end{tabular}
\end{ruledtabular}
\end{table}

To confirm the monoclinic symmetry of our single crystals, we performed single-crystal X-ray diffraction at room temperature using a Bruker APEX II, a standard kappa-diffractometer. This system is equipped with a molybdenum X-ray tube and utilizes monochromatic Mo-K$\alpha$ radiation ($\lambda$ = 0.70930 \AA) and an Apex2 area detector. Data reduction was performed using the CrysAlisPro package program from Oxford Diffraction~\cite{crysalis2014agilent}. 
The crystal exhibited twinning along the c axis with a rotation of 119.85$\degree$ between the twin components. 
97$\%$ of all measured reflections could be attributed to a single structural domain, while the remaining 3$\%$ indicate the presence of a secondary twin domain. It is important to note that the number of assigned reflections does not allow for a quantitative determination of the domain volume fractions. Nevertheless, the fact that nearly all reflections can be assigned to one domain strongly suggests that it is predominant, which is fully consistent with the results of our other measurements. 

In Fig.~\ref{fig:xray}, we present the structure factor along the $(h k 0)$ and $(0 k l)$ planes in reciprocal space. The vertices of the dashed lines in the figures indicate the sites of the reciprocal lattice in the monoclinic phase, which coincide perfectly with all reflections originating from the sample. Further, the single-crystal diffraction pattern of Na$_3$Co$_2$SbO$_6$ was successfully indexed to a centered monoclinic cell in the space group $C2/m$ (No.\ 12) using the Jana2020 software \cite{Jana2020}. The structural parameters obtained through the Rietveld method are presented in Table~\ref{tab:xray-refinement}.

The crystal structure is shown in Fig.~\figcrystalstructure(a) of the main text. As it can be seen, within the magnetoactive layer, Na$_3$Co$_2$SbO$_6$ features a honeycomb arrangement characterized by edge-sharing CoO$_6$ octahedra, with an SbO$_6$ octahedron located at the center of each honeycomb. These honeycomb layers are aligned along the crystallographic $c$ axis and are separated by a nonmagnetic layer of Na$^+$ ions.

\section{Magnetization and magnetocaloric measurements}

\begin{figure*}[tb!]
\includegraphics[width=1\textwidth]{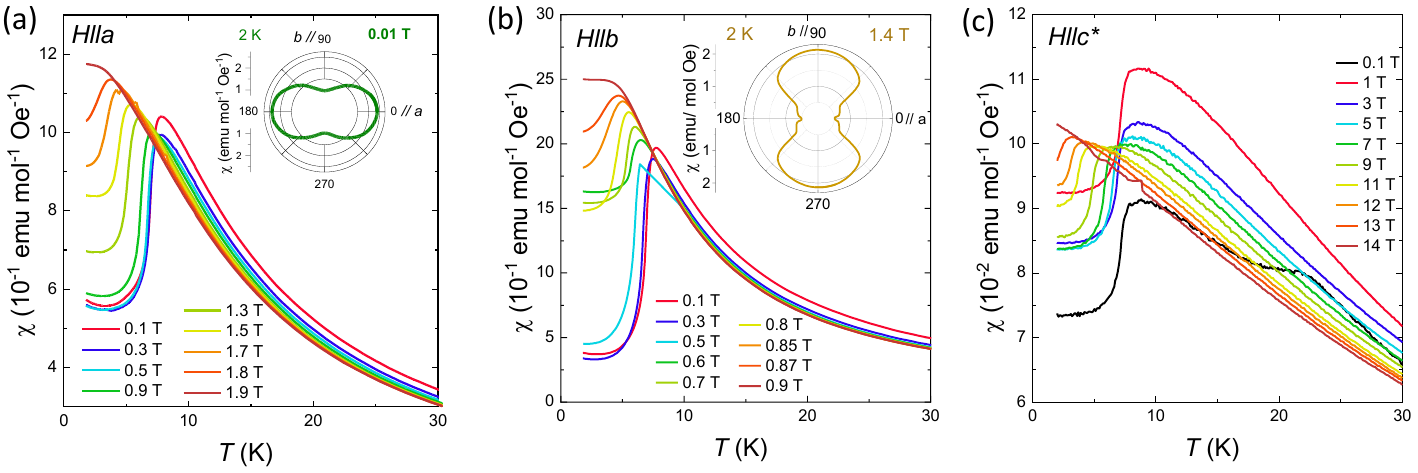}
\caption{%
(a)~Magnetic susceptibility $\chi$ as a function of temperature $T$ for different fields $\mathbf H \parallel \mathbf a$.
(b)~Same as (a), but for $\mathbf H \parallel \mathbf b$. The insets in (a) and (b) shows the in-plane angular dependence of the magnetic susceptibility for $\mu_0 H$ = 0.01 T and 1.4 T, respectively.
(c)~Same as (a), but for $\mathbf H \parallel \mathbf c^*$. The broad anomaly at $\sim 20$\,K indicates the onset of short-range ferromagnetic chains.}
\label{fig:MT}
\end{figure*}

\begin{figure*}[tb!]
\includegraphics[width=1\textwidth]{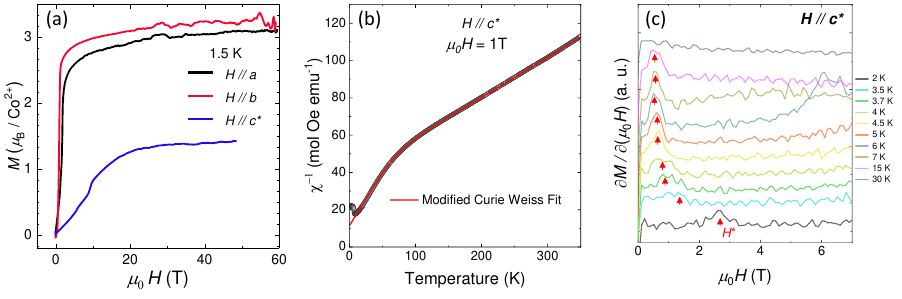}
\caption{%
(a)~High-field isothermal magnetization $M$ as a function of field $H$ at $T = 1.5$\,K measured in pulsed magnetic fields along the $a$, $b$, and $c^*$ directions.
(b)~Inverse of the magnetic susceptibility $\chi^{-1}$ as a function fo temperature $T$ for $\mathbf H \parallel \mathbf c^*$ at $\mu_0 H = 1$\,T. The red line represents a modified Curie-Weiss fit, accounting for both the ground state and the first excited state, separated by an excitation gap. (c) First derivative of the isothermal magnetization along $c^*$ direction at selected temperatures. $H^*$ represents the low-field anomaly from the short-range ferromagnetic chains.
}
\label{fig:pulse field MH and modified CW fit}
\end{figure*}

\begin{figure*}[tb!]
\includegraphics[width=1\textwidth]{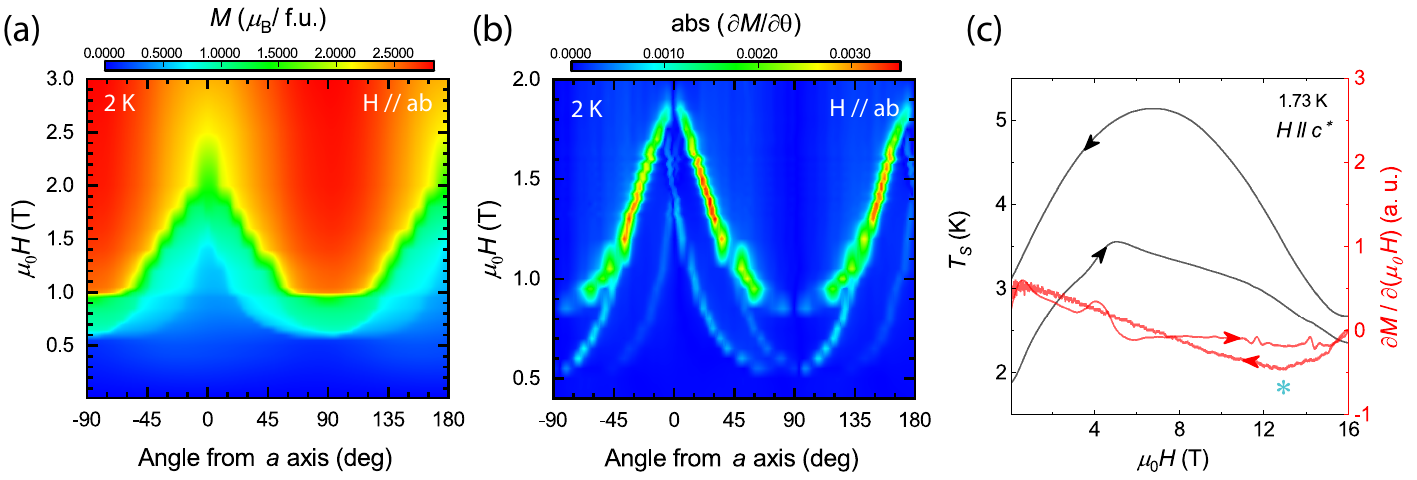}
\caption{ Angular dependent in-plane (a) magnetization and (b)~ angle derivative of magnetization at 2 K. The angular dependent magnetisation is performed in a field-up sequence.
(c)~Sample temperature $T_\text{S}$ (black) and its derivative $\partial T_\text{S}/\partial (\mu_0H)$ as a function of field $\mathbf H  \parallel \mathbf c^*$ from low-temperature MCE measurements in pulsed fields under quasi-adiabatic conditions. Arrows indicate the field-up and field-down sweeps, while asterisks mark the field-induced transitions.}
\label{fig:angulardependence}
\end{figure*}

In this supplemental section, we present details of the DC magnetic studies conducted in both static and pulsed fields, along with additional analysis, and further magnetocaloric effect data (MCE).
Figure~\ref{fig:MT} shows the temperature dependence of the DC magnetic susceptibility $\chi$ with applied magnetic fields along the crystallographic $a$, $b$ and $c^*$ directions. In the magnetic long-range ordered state below $T_\textrm{N} = 7$\,K, the magnetic susceptibility decreases for all three directions. Our samples exhibit a well-defined phase transition with minimal variation across different samples, suggesting a homogeneous distribution of Na close to ideal stoichiometry. The angular dependence of the magnetic susceptibility at $T$ = 2\,K with a magnetic field of 0.01\,T and 1.4\,T applied in the $ab$ plane is shown in the insets of Fig.~\ref{fig:MT}(a) and (b), respectively. At low fields, it exhibits a two-fold symmetry indicative of the broken $C_3$ symmetry. Furthermore, at high magnetic fields the magnetic anisotropy deviates from the expected sinusoidal behavior for $C_2$ symmetry (see the inset of Fig.~\ref{fig:MT}(b)). Overall, $T_\textrm{N}$ continuously decreases with the increase of the magnetic field and is fully suppressed at 1.8\,T, 0.87\,T and 13\,T for the $a$, $b$ and $c^*$ directions, respectively. We observe a slightly higher critical field $H_{a2}$ from magnetic susceptibility compared to the value of 1.68 T obtained from isothermal magnetization at 2 K for $\mathbf{H} \parallel \mathbf{a}$, which we attribute to a slight sample misorientation($\sim 10^o$).

Figure~\ref{fig:pulse field MH and modified CW fit}(a) shows the pulsed-field isothermal magnetization at $T = 1.5$\,K as a function of field along the three crystallographic directions, up to 60\,T. Note that the data shown have been corrected by subtracting the Van Vleck contribution for Co$^{2+}$ ions, which is $0.01\,\mathrm{T}/\mu_\mathrm{B}$~\cite{Zhang2024NCTO}.

The excitation gap between the $j_\text{eff}$ = 1/2 and $j_\text{eff}$ = 3/2 states was estimated by fitting the experimental susceptibility data shown in Fig.~\ref{fig:pulse field MH and modified CW fit}(c) using a modified Curie-Weiss expression~\cite{Besara2014gap},
\begin{equation}
\chi (T) = \chi_0 + \frac{1}{T-\Theta_\textrm{W}} \frac{N_\textrm{A}}{3k_\textrm{B}} 
\left( \frac{\mu^2_{0, \textrm{eff}} + \mu^2_{1, \textrm{eff}} \textrm{e}^{-\frac{\Delta}{T}}}
{1 + \mu^2_{1, \textrm{eff}} e^{-\frac{\Delta}{T}}} \right),
\label{eqn:modified curie weiss fit}
\end{equation}
with the Boltzmann constant $k_B$, the Avogadro constant $N_A$, the Curie-Weiss temperature $\Theta_\mathrm{W}$, the excitation gap $\Delta$, the ground-state effective magnetic moment $\mu_{0,\textrm{eff}}$, and the first-excited state effective magnetic moment $\mu_{1,\textrm{eff}}$. 
The fitting yields $\chi_0 = 0.0064(2)$\,emu/(mol Oe), $\Delta = 247(3)$\,K, $\Theta_\mathrm{W} = -14.2(9)$\,K, $\mu_{0, \textrm{eff}}/\mu_\mathrm{B} = 3.52(5)$, and $\mu_{1, \textrm{eff}}/\mu_\mathrm{B} = 6.69(11)$. The observed $\mu_{0, \textrm{eff}}/\mu_\mathrm{B}$ is close to the expected value of $3.75$ for a $j_{\textrm{eff}} = \frac{1}{2}$ state with an isotropic Land\'e $g$-factor $g_\text{iso} = 4.33$ neglecting the anisotropy for simplicity.
%
%
This suggests that the ground state of Na$_3$Co$2$SbO$6$ is characterized by $j_{\textrm{eff}} = \frac{1}{2}$, while the first excited state corresponds to $j_{\textrm{eff}} = \frac{3}{2}$, with an excitation energy of approximately 247\,K.

Fig.~\ref{fig:pulse field MH and modified CW fit} (c) depicts the first derivative of the isothermal magnetization along the $c^*$ direction at selected temperatures between 2 and 30\,K. The broad anomaly at $H^*$ represents the low-field anomaly from the short-range ferromagnetic chains persisting above $T_\textrm{N}$ up to about 20\,K.

Figures~\ref{fig:angulardependence}(a) and \ref{fig:angulardependence}(b) show the angular-dependent isothermal magnetization and its field-angle derivative, respectively, at 2 K for a magnetic field applied in the $\textit{ab}$ plane. The hysteretic critical fields $H_{a1}$ and $H_{b1}$ along the $a$ and $b$ directions, respectively, each split into two when the field is applied in-plane but slightly away from these principal directions, suggesting the presence of a possible field-induced tricritical point~\cite{Mi2025}. The angular magnetization exhibits a steep field-angle gradient near the $\mathbf{a}$ direction and a more gradual gradient near the $\mathbf{b}$ direction, resulting in sharp and gentle changes in the critical fields in the vicinity of the $\mathbf{a}$ and $\mathbf{b}$ directions, respectively.

Figure~\ref{fig:angulardependence}(c) presents the magnetocaloric effect and its first derivative for $\mathbf{H} \parallel \mathbf{c}^*$ in pulsed magnetic fields under quasi-adiabatic conditions. A change in slope at $\mu_0H_{c2} = 12.8(8)$\,T in the first derivative marks the onset of the field-polarized state. No anomaly is observed at $\mu_0H_{c1}$, likely due to the broad nature of the MCE around this critical field. The small anomaly near $\sim 4$\,T is an artifact.

\section{Specific-heat measurements}

\begin{figure*}[tb!]
\includegraphics[width=1\textwidth]{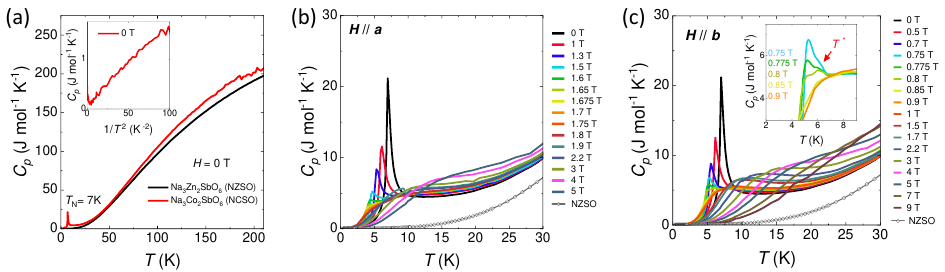}
\caption{%
(a)~Specific heat at constant pressure $C_p$ as a function of temperature $T$ for  single-crystal Na$_3$Co$_2$SbO$_6$ and pelletized polycrystalline Na$_3$Zn$_2$SbO$_6$ at zero field. The inset shows $C_p$ versus $1/T^2$ for Na$_3$Co$_2$SbO$_6$ at temperatures down to 100\,mK, highlighting a nuclear Schottky contribution.
(b)~Same as (a), but with an applied magnetic field along the $a$ axis.
(c)~Same as (a), but with an applied magnetic field along the $b$ axis. The inset highlights the emergence of short-range order at $T^*$, as indicated by the red arrow.}
\label{fig:specific heat}
\end{figure*}

In this supplemental section, we provide details of the specific heat measurements for Na$_3$Co$_2$SbO$_6$ and its nonmagnetic analogue, Na$_3$Zn$_2$SbO$_6$.
Figure~\ref{fig:specific heat}(a) illustrates the temperature evolution of the zero-field specific heat $C_p$ of our Na$_3$Co$_2$SbO$_6$ single crystal and its nonmagnetic analogue Na$_3$Zn$_2$SbO$_6$ pelletized polycrystal. The specific heat of Na$_3$Co$_2$SbO$_6$ exhibits a clear lambda-type anomaly at $T_\textrm{N} = 7.0(1)$\,K, indicating the onset of long-range magnetic order. At temperatures below 1\,K, $C_p$ exhibits a Schottky tail, $C_p \propto 1/T^2$, arising from the hyperfine interaction between the electron and nuclear spins $I = \frac{7}{2}$ of Co$^{2+}$ ions~\cite{Tari2003}, as shown by the linear slope in the $C_p$ versus $1/T^2$ plot in the inset of Fig.~\ref{fig:specific heat}(a). The phonon contribution to the specific heat of Na$_3$Co$_2$SbO$_6$ was obtained by scaling the specific heat of Na$_3$Zn$_2$SbO$_6$ using the Lindemann scaling factor $\theta_D^{\text{NCSO}} / \theta_D^{\text{NZSO}} = 1.016$, accounting for differences in molecular mass and molar volume between the two compounds~\cite{Tari2003}. 

Figures~\ref{fig:specific heat}(b) and \ref{fig:specific heat}(c) display the specific heat curves $C_p(T)$ with applied magnetic fields along the two in-plane directions.
With the increase of the magnetic field, $T_\textrm{N}$ decreases and the anomaly is completely suppressed at 1.65\,T and 0.85\,T for the $a$ and $b$ direction, respectively, marking the complete suppression of the magnetic order.
For fields $\mathbf H \parallel \mathbf b$ between 0.7\,T and 0.85\,T, a shoulder-like peak was observed at a temperature $T^*$, indicated in the inset of Fig.~\ref{fig:specific heat}(c). Its shape suggests the onset of short-range magnetic order, in line with the magnetization analysis presented in the main text. The broad maximum observed in the field-polarized state results from the opening of a spin-excitation gap, which was fitted to an exponential behavior in the low-temperature regime for the in-plane field directions. The resulting spin-excitation gaps are shown in the insets of Figs.~\figphasediag(a) and \figphasediag(b) in the main text. Note that specific-heat measurements for the $c^*$ direction were not possible due to the large magnetic torque, which would damage the sample holder (puck) in our PPMS device.

\section{Classical Monte Carlo simulations}

\begin{figure*}[tb!]
\includegraphics[width=\linewidth]{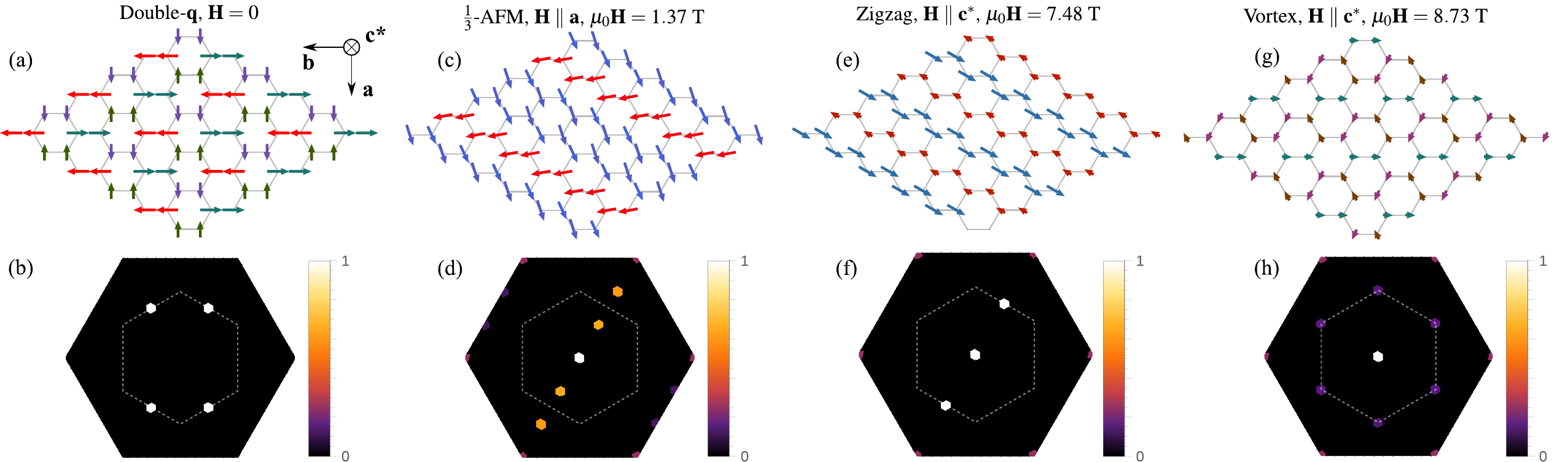}
\caption{%
(a)~Representative spin configuration of the double-$\mathbf q$ ground state at $T=0$ from classical Monte Carlo simulations of the extended Kitaev-Heisenberg model defined in the main text at vanishing field $\mathbf H = 0$. Arrows indicate spin directions projected onto the $ab$ plane.
(b)~Static spin structure factor of the single configuration shown in (a), displaying two Bragg peaks at the centers $\mathbf{M}_1$ and $\mathbf{M}_2$ of two edges of the first Brillouin zone (gray dashed hexagon).
(c)~Same as (a), but for the 1/3-AFM ground state for $\mathbf{H}\parallel\mathbf{a}$ at $\mu_0 \mathbf H = 1.37$ T.
(d)~Same as (b), but for the 1/3-AFM ground state for $\mathbf{H}\parallel\mathbf{a}$ at $\mu_0 \mathbf H = 1.37$ T, with Bragg peaks at $\frac{2}{3}\mathbf{M}_1$ and $\frac{4}{3}\mathbf{M}_1$.
(e)~Same as (a), but for the zigzag ground state for $\mathbf{H}\parallel\mathbf{c}^*$ at $\mu_0 \mathbf H = 7.48$ T.
(f)~Same as (b), but for the zigzag ground state for $\mathbf{H}\parallel\mathbf{c}^*$ at $\mu_0 \mathbf H = 7.48$ T, with a Bragg peak at $\mathbf M_2$.
(g)~Same as (a), but for the vortex ground state for $\mathbf{H}\parallel\mathbf{c}^*$ at $\mu_0 \mathbf H = 8.73$ T.
(h)~Same as (b), but for the vortex ground state for $\mathbf{H}\parallel\mathbf{c}^*$ at $\mu_0 \mathbf H = 8.73$ T, with Bragg peaks at the corners $\mathbf K$ and $\mathbf K'$ of the first Brillouin zone.
The Bragg peaks at $\boldsymbol{\Gamma}$ in the static structure factors in (d, f, h) reflect the canted nature of the state, indicating a finite magnetization along the field direction.
}
\label{fig:theorysupplgroundstates}
\end{figure*}

\begin{figure*}[tb!]
\includegraphics[width=\linewidth]{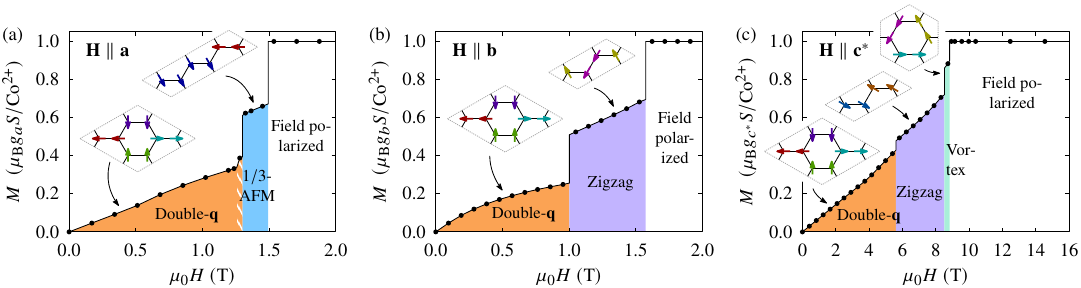}
\caption{%
(a)~Magnetization $M$ as a function of in-plane field $\mathbf H \parallel \mathbf a$ at $T=0$ from classical Monte Carlo simulations of the extended Kitaev-Heisenberg model defined in the main text. Insets show spin directions projected onto the $ab$ plane within the magnetic unit cell for each commensurate phase.
(b)~Same as (a), but for the other in-plane field direction $\mathbf H \parallel \mathbf b$.
(c)~Same as (a), but for the out-of-plane field direction $\mathbf H \parallel \mathbf c^*$.}
\label{fig:theory-m-vs-h}
\end{figure*}

In this supplemental section, we present details of the Monte Carlo simulations and further characterization of the field-induced phases.
The model in Eq.~(\eqtheorymodel) of the main text is defined on a honeycomb lattice with $L\times L$ unit cells, each consisting of two sites, with periodic boundary conditions. In the classical limit, the spins are considered as three-component vectors of unit length $\mathbf{S}_i^2=1$. Both local Metropolis and overrelaxation algorithms are employed to update the spin configurations. Additionally, parallel tempering is incorporated to enhance equilibration of replicas at low temperatures; see Ref.~\cite{francini2024ferrimagnetism} for further details.
The classical ground-state configurations shown in Figs.~\figphasediag(d)--(f) of the main text are obtained using an iterative minimization algorithm applied to the lowest-temperature replica of the parallel-tempering Monte Carlo history. Details of the minimization algorithm are described in Ref.~\cite{janssen2016heisenbergkitaev}.
The system converges to different ground states depending on the strength and direction of the external field.
As the magnetic field $\mathbf{H}$ increases, the system passes through a sequence of distinct field-induced ordered phases before reaching the fully polarized state:
double-$\mathbf{q}$, 
and 1/3-AFM for $\mathbf{H} \parallel \mathbf{a}$, 
double-$\mathbf{q}$ and canted zigzag for $\mathbf{H} \parallel \mathbf{b}$, 
and double-$\mathbf{q}$, canted zigzag, and vortex phases for $\mathbf{H} \parallel \mathbf{c}^*$.
The in-plane spin configurations and the corresponding static spin structure factors of the four commensurate field-induced phases are illustrated in Fig.~\ref{fig:theorysupplgroundstates}.
Here, the spin projections onto the $ab$ plane are shown, such that shorter arrows indicate smaller in-plane components and larger out-of-plane $c^*$ components.
Note that for $\mathbf{H} \neq 0$, the presence of a peak at the $\Gamma$ point in the spin structure factor indicates a finite magnetization induced by the external field.
Figure~\ref{fig:theory-m-vs-h} shows the magnetization as a function of field $\mathbf H$ for the three different field directions.
The corresponding low-temperature phase diagrams are compared with the experimental results in Fig.~\figphasediag\ of the main text.
We note that in the classical simulations, the transitions from the field-induced intermediate phases to the polarized state are strongly first order for fields applied along the in-plane $a$ and $b$ directions. However, the magnetization jump is expected to significantly decrease with the inclusion of quantum fluctuations and may eventually vanish in the spin-1/2 case~\cite{janssen2016heisenbergkitaev, janssen17}.

\bibliographystyle{longapsrev4-2}
\bibliography{NCSO}